\documentclass{article}
\usepackage{amsmath,graphicx,color}
\usepackage[T1]{fontenc} % add special characters (e.g., umlaute)
\usepackage[utf8]{inputenc} % set utf-8 as default input encoding
\usepackage{cite,url}
\usepackage{xcolor}
\usepackage{booktabs}
\usepackage{enumitem}
\usepackage{amsfonts}
\usepackage{dsfont} 
\usepackage{multirow}
\usepackage{comment}
\usepackage{makecell} 
\usepackage{nicefrac}
\usepackage[bookmarks=false,hidelinks]{hyperref}
\usepackage{lineno}
\usepackage[preprint]{spconf}

\def\mycopyrightnotice{%
  \begin{minipage}{\textwidth}
 \scriptsize
  \copyright~2023 IEEE. Personal use of this material is permitted. Permission from IEEE must be obtained for all other uses, in any current or future media, including reprinting/republishing this material for advertising or promotional purposes, creating new collective works, for resale or redistribution to servers or lists, or reuse of any copyrighted component of this work in other works.
  \end{minipage}
}

\title{On the effect of data-augmentation on local embedding properties in the contrastive learning of music audio representations}

\name{Matthew C. McCallum, Matthew E. P. Davies, Florian Henkel, Jaehun Kim, Samuel E. Sandberg}
\address{SiriusXM-Pandora, USA}

\begin{document}
\ninept
\maketitle
\begin{abstract}
Audio embeddings are crucial tools in understanding large catalogs of music. Typically embeddings are evaluated on the basis of the performance they provide in a wide range of downstream tasks, however few studies have investigated the local properties of the embedding spaces themselves which are important in nearest neighbor algorithms, commonly used in music search and recommendation. In this work we show that when learning audio representations on music datasets via contrastive learning, musical properties that are typically homogeneous within a track (e.g., key and tempo) are reflected in the locality of neighborhoods in the resulting embedding space. By applying appropriate data augmentation strategies, localisation of such properties can not only be reduced but the localisation of other attributes is increased. For example, locality of features such as pitch and tempo that are less relevant to non-expert listeners, may be mitigated while improving the locality of more salient features such as genre and mood, achieving state-of-the-art performance in nearest neighbor retrieval accuracy. Similarly, we show that the optimal selection of data augmentation strategies for contrastive learning of music audio embeddings is dependent on the downstream task, highlighting this as an important embedding design decision.
\end{abstract}

\copyrightnotice{\mycopyrightnotice}

\section{Introduction}
\label{sec:introduction}

\subsection{Overview}
\label{subsec:overview}

For industrial-scale music catalogs, audio embeddings are crucial in estimating perceptual and musical similarity between audio waveforms, and further, categorizing them into any taxonomy of human-readable labels. The dimensionality reduction from audio waveform to embedding ensures tasks may be addressed efficiently in terms of downstream model complexity, and pairwise embedding comparison. The generality of embeddings provides further computational efficiency in that large music catalogs may be analyzed just once, and thereafter adapted to any number of labeling tasks.

Recently, benefiting from the ability to learn from large amounts of unlabelled data, self-supervised contrastive audio embeddings have been shown to establish new state-of-the-art (SotA) results on a number of downstream tasks. Many use little to no data augmentation. COLA \cite{saeed2021contrastive} established a new SotA on tasks concerning speech, music and environmental noise, by sampling positive pairs as distinct samples from audio regions of 10 seconds in length. \cite{wang2022towards}, which employed only mix-up augmentation \cite{zhang2017mixup}, further improved on many of these results. Other contrastive music embedding models are also trained without data-augmentation \cite{alonso2022music, alonso2023pre}. 

In many cases data augmentation may not be necessary to increase the total coverage of music audio properties (e.g., pitch, tempo, production, etc.) in the dataset, as typically the datasets employed are large enough to include this variability naturally. However, this variability may not be seen within contrastive pairs. For example, the strategy of sampling pairs from local regions (e.g., $\pm 10$~s) of audio waveforms employed in these previous works likely causes low variability of certain properties such as key, tempo and equalization, within contrastive positive pairs. Due to this intra-pair homogeneity, we hypothesize that the local regions of the resulting embedding space will be similarly homogeneous. Yet, little research (e.g., \cite{DBLP:journals/fams/KimULH19}) has investigated the local properties of audio embedding spaces, and how audio characteristics are reflected therein. While performance on certain tasks such as pitch and key estimation proves the existence of pitch information within contrastively learned embeddings \cite{mccallum22ismir, wang2022towards}, this might not necessarily guarantee the reliable emergence of the information within a local vicinity in the embedding space. The locality of this information in that space is important because such local properties are directly exposed in approximate nearest neighbor (ANN) algorithms which play important roles in efficient audio file identification, search, and recommendation, where the nearest neighbors of an audio query will have similarities determined by the local properties of the embedding space~\cite{DBLP:journals/pr/MinLH04}.

Local organization or manipulation of an embedding space by tempo \cite{mccallum24icassp}, pitch, or key may be desirable, e.g., for professional music production or DJ applications where they are important. However, it may be problematic for consumer music applications where pitch, key and tempo can be less relevant to listeners than properties such as genre and instrumentation, that are in many cases only loosely correlated with the former. When the organization of embedding spaces by such properties is undesirable, we hypothesize that data augmentation strategies that increase the intra-pair variability of key, tempo and equalization, can reduce the sensitivity of embeddings to such properties relative to others, e.g., genre and instrumentation. Thus we expect the resulting embeddings to be more useful in nearest neighbor and labeling tasks concerning the latter.

\subsection{Related Work}
\label{subsec:relatedwork}

Various approximations of pitch shifting \cite{spijkervet2021contrastive, kharitonov2021data, choi2022towards, niizumi2022byol, wang2021multi, yao2022contrastive, zhao2022s3t, jansen2018unsupervised}, time stretching \cite{zhao2022s3t, niizumi2022byol, choi2022towards}, and equalization \cite{spijkervet2021contrastive, kharitonov2021data, yao2022contrastive} have been employed in contrastive learning. We note that time-stretching in the learning of contrastive music representations is particularly rare, with only \cite{zhao2022s3t, choi2022towards} employing this augmentation, and no results specifically demonstrating its efficacy. Furthermore, few studies investigate the efficacy of audio music embeddings in temporal downstream tasks such as beat/tempo estimation \cite{li2023mert}. However, we find the effect of time-stretching augmentation to be particularly salient in this research.

In many studies data augmentation strategies are chosen based on their computational efficiency, and their suitability for the media at hand (e.g., for music, speech or environmental noise). Often data augmentation is decided on this basis prior to experimentation and left unchanged, although some studies performed ablation demonstrating some effects of data augmentation. For example, \cite{spijkervet2021contrastive} studied the effect of individual augmentations on the Magnatagatune dataset \cite{law2009evaluation}, highlighting equalization as important in the contrastive learning of music embeddings from time domain audio, for this task. \cite{kharitonov2021data} found pitch shifting to be particularly important in the learning of speech representations. \cite{wang2021multi} studied the effect of pairs of augmentations on the Audioset dataset highlighting a combination of pitch shifting and mix-up to be effective.

While this work investigates the effect of augmentation on local embedding space properties, other studies have proposed designing supervised embedding spaces by training subspaces on specific concepts (e.g., genre, mood, era, instruments) via disentangled metric learning \cite{lee2020metric, lee2020disentangled}. These subspaces may be weighted to emphasize certain properties in downstream tasks, however, no results were provided as to the independence of each of these subspaces.

Due to the generality of audio embeddings and the number of possible downstream applications, this wide body of literature provides a patchwork of approaches to designing them for various downstream tasks. However, there are two studies we see missing from the literature. First, while there are some results demonstrating the effects of data augmentation on particular downstream tasks there are no results demonstrating the effect of data augmentation on local properties of the embedding space which are important for efficient ANN solutions employed in search and recommendation. Secondly, ablation studies on data augmentation for audio embeddings are based on the results from a single task, however, we expect data augmentation strategies to have task-dependent effects. In this work we address these areas by first demonstrating the effect of data augmentation on local embedding properties, and secondly, we show the subsequent effect of various data augmentation methods on downstream nearest neighbor and labelling tasks. We highlight time-stretching as an important and under-explored augmentation and achieve SotA performance on a number of tasks.

\section{Methodology}
\label{sec:pagestyle}

\begin{figure}[t!]
 \centerline{
 \includegraphics[trim={0 0 0 0},clip,width=0.8\columnwidth]{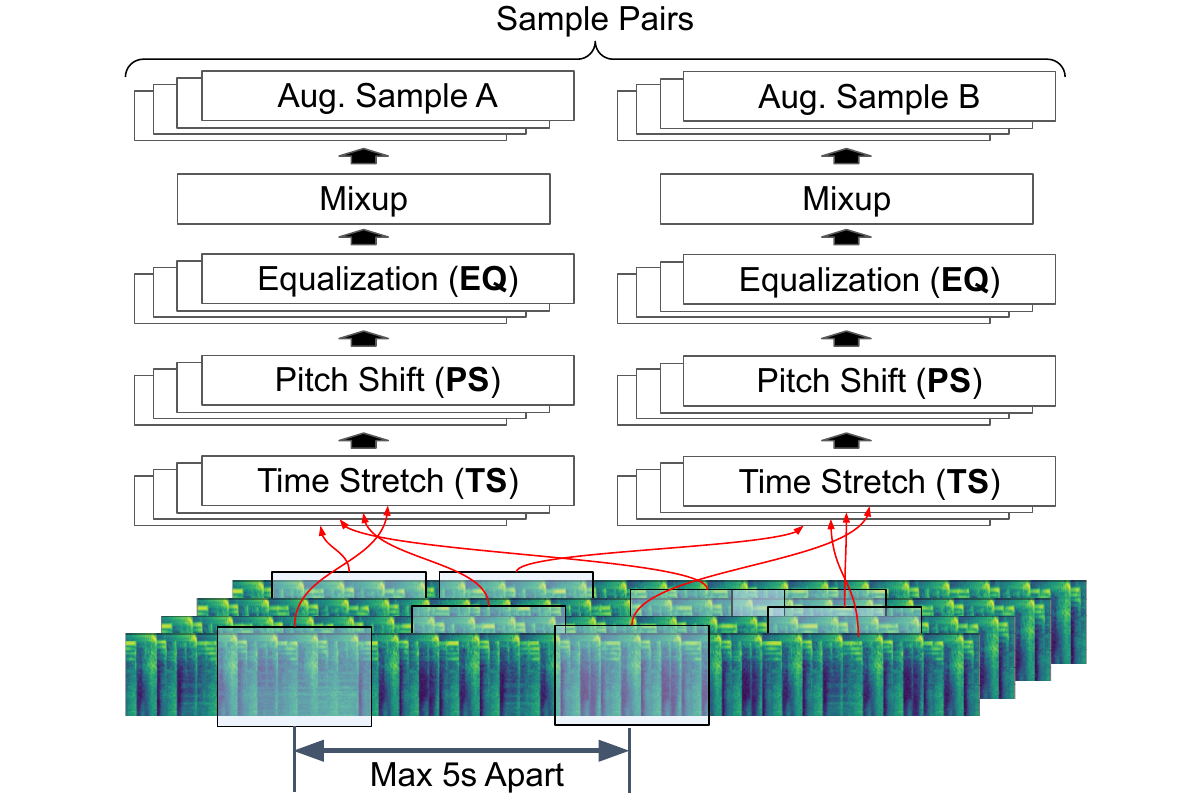}}
 \caption{Sampling and augmentation pipeline diagram.}
 \label{fig:pipeline}
\end{figure}

Here we study the open-source Musicset-Unsupervised Large Embedding (MULE) model of \cite{mccallum22ismir} which employs a convolutional architecture first proposed in \cite{wang2022towards}. This provides a reproducible and performant baseline for self-supervised contrastive learning of music audio. It is trained via a familiar strategy of sampling contrastive pairs of mel-spectrograms that are located locally ($\pm 5$~s) in music audio timelines. A similar strategy is common in the literature \cite{spijkervet2021contrastive, wang2022towards, mccallum22ismir, choi2022towards, alonso2022music, alonso2023pre, saeed2021contrastive, mccallum2019unsupervised} and has been shown to establish a new SotA in self-supervised representations for a number of tasks \cite{mccallum22ismir, wang2022towards}. NT-XEnt loss is employed \cite{chen2020simple} where the normalized temperature-scaled crossentropy of the cosine distance between positive pairs is minimized relative to all other examples in a given batch, which form the negative samples. Based on this loss, it is expected that characteristics that are typically common in positive pairs would be embedded locally in an embedding space with respect to cosine distance. Here we investigate three such properties: tempo, key and equalization. We employ a pipeline implementing the related data augmentation strategies of time-stretching, pitch-shifting and equalization, shown in Fig.~\ref{fig:pipeline}. Specifically, we augment mel spectrograms of the form,

\begin{equation}
    \resizebox{0.90\hsize}{!}{
$X[u,m] = \log_{10}\left(\sum\limits_{k=0}^{k=K} S_u[k] \left\vert \sum\limits_{n=0}^{n=N}{e^{-\frac{2 \pi n k j}{K}}w[n]x[ml-n]} \right\vert \right)$ .
}
\end{equation}
$S_u[k]$ for $0\leq u < U$ is a mel window at index $u$ according to HTK mel scaling \cite{young2002htk}. $w[n]$ a windowing function of length $N$, $l$ the hop size, and $K$ the DFT size. Parameters used here are identical to \cite{mccallum22ismir}.

\textbf{Time stretching augmentation} (\textbf{TS}), defined as $X_{TS}[u,m]=h_{TS}(X[u,m]; \tau)$, is applied by resampling $X[u,m]$ using cubic spline interpolation at points $t=\tau m$. We sample $\tau$ according to $\mathbf{\tau} \sim 1 / (\tau \log{(1.5/0.75)})$ in range $[0.75,1.5]$. To allow for $\tau>1.0$, $X[u,m]$ is sampled with a context of $4.5$ seconds and then truncated to $3$ seconds following augmentation. While time-stretching is seldom applied in music representation learning \cite{zhao2022s3t, choi2022towards}, we note that time-warping, where a randomly sampled $m$ will be shifted in time by resampling strategies was employed in \cite{park2019specaugment, zhao2022s3t}. Further, one effect of the Random Resized Crop (RRC) augmentation of \cite{niizumi2022byol, nguyen2023improving, ghosh2022mast, fonseca2021unsupervised} 
could be considered a time stretching augmentation, albeit coupled with a similar augmentation on the frequency axis.

\textbf{Pitch shifting augmentation} (\textbf{PS}), defined as $X_{PS}[u,m]=h_{PS}(X[u,m]; \mu)$, is applied via cubic spline interpolation at points:
\begin{equation}\label{eq:pitch_shift}
f = S_U \log_{10}(1+\mu(10^{u/S_U} - 1)),
\end{equation}
with $S_U=U/\log_{10}(1+R/700)$ where $R$ is the sampling rate in Hz. This is equivalent to interpolating at points $\hat{f}=\mu \hat{m}$ on a linear frequency scale then translating to HTK mel scaling. $\mu$ is distributed as $\mathbf{\mu} \sim 1 / (\mu \log{(1.335/0.749)})$ in the range $[0.749,1.335]$, corresponding to a relative decrease / increase of up to 5 semitones. When $\mu<1.0$, interpolated frequency bins are set to $0.0$ for $u>\mu U$. We note that the pitch augmentation here has the imperfect effect of changing the bandwidth of harmonic partials,\footnote{Future work may use constant-Q features, thereby removing this artifact.} yet is computationally efficient and a more accurate pitch shift operation than the translation operation of \cite{wang2021multi, jansen2018unsupervised}, or the RRC augmentation in \cite{niizumi2022byol, nguyen2023improving, ghosh2022mast, fonseca2021unsupervised}.

\textbf{Equalization augmentation} (\textbf{EQ}) is achieved by applying randomly sampled lowpass and highpass filters to features,
\begin{equation}\label{eq:eq}
X_{EQ}[u,m] = X[u,m] + \log_{10}\left(\sum\limits_{k=0}^{k=K} S_u[k] B_{f, c}[k] \right) ,
\end{equation}
where $B_{f, c}[k]$ refers to a third-order butterworth magnitude response, with corner frequency $f$ and highpass / lowpass selection $c\in \left\{ 0, 1\right\}$. We follow \cite{spijkervet2021contrastive} where center frequencies are uniformly sampled from the range 2.2~kHz to 4~kHz for lowpass, and the range 200~Hz to 1.2~kHz for highpass filters. For each sample, we apply equal probabilities of highpass, lowpass and no filter.

In addition to \textbf{TS}, \textbf{PS}, \textbf{EQ}, we investigate the baseline \textbf{RRC}, and the combinations \textbf{TSPSEQ} and \textbf{TSPS} according to the order of operations in Fig.~\ref{fig:pipeline}. The latter is similar to RRC, but with a more accurate pitch-shift operation. For each augmentation strategy, or combination thereof, we fine-tune all MULE layers using batches of $960$ example pairs, sampled and augmented in real-time from full track length timelines of a dataset of $1.7$M music tracks. In all cases we train for $100$k steps with a learning rate linearly ramping from $0$ to $0.001$ over $5$k steps followed by a cosine decay to $0$ over the remaining $95$k steps. All other hyperparameters are identical to \cite{mccallum22ismir}.

\section{Results}
\label{sec:results}

We consider three key categories of datasets -- tempo, pitch and mixed labelling. For tempo, we consider the same collection of datasets in \cite{boeck20ismir} plus the Harmonix dataset \cite{nieto19ismir}. We hold out Gtzan \cite{marchand2015swing}, Giantsteps-Tempo \cite{schreiber2018crowdsourced, knees2015two}, and ACM-Mirum \cite{percival2014streamlined} as individual test sets, using the most recent annotations for each. Collectively, we refer to these splits as the AllTempo dataset. Other datasets employed include NSynth pitch and instrument datasets (NSynth\textsubscript{P}, NSynth\textsubscript{I}) \cite{engel2017neural}, Giantsteps Key (GS\textsubscript{Key}) \cite{knees2015two}, Gtzan \cite{tzanetakis2002musical}, Magnatagatune (MTT) \cite{law2009evaluation}, and the Million Song Dataset (MSD) \cite{bertin2011million} (employing the commonly used variation of the 50 most common labels.)\footnote{\url{https://github.com/jongpillee/music_dataset_split}} NSynth\textsubscript{P} and GS\textsubscript{Key} refer specifically to pitch-related information. In all cases we use the same partitions and filtering as in \cite{mccallum22ismir}.

\subsection{Local Embedding Properties}
\label{subsec:localembeddingproperties}

\begin{figure}[t!]
 \centerline{
 \includegraphics[trim={0.1cm 0.35cm 1.7cm 1.3cm},clip,width=1\columnwidth]{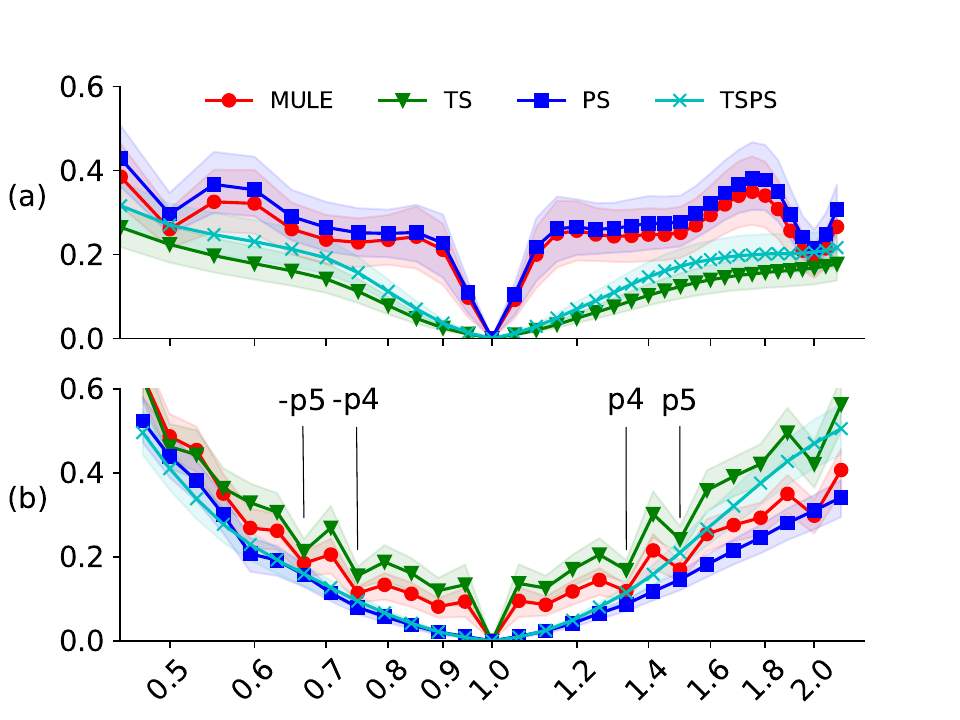}}
 \caption{Mean and interquartile range of cosine distance between embeddings of unmodified tracks and (a) time-stretched tracks, and (b) pitch-shifted tracks, for different augmentation pipelines.}
 \label{fig:manipulated_tempo}
\end{figure}

To demonstrate the effects of tempo / pitch augmentation on embedding spaces, we synthetically time-stretch all tracks in all AllTempo test partitions, and pitch-shift all tracks in the GS\textsubscript{Key} dataset using Sox.\footnote{https://sox.sourceforge.net/} For each stretch / shift factor we measure the cosine distance between the track-average embeddings of the modified and unmodified versions. The mean and interquartile range across the respective datasets is shown in Fig.~\ref{fig:manipulated_tempo}. For models trained without TS augmentation we observe musically sensical troughs in cosine distance at half and double time. Similarly, we observe that models trained without PS augmentation show musically sensical troughs in cosine distance at perfect 4th, perfect 5th and octave intervals, and peaks at minor 2nd and diminished 5th intervals. For each property the corresponding data augmentation removes both these musical properties, and mitigates the movement of the modified embeddings away from their unmodified version. In both cases PS augmentation increases embedding sensitivity to tempo, whilst TS augmentation increases embedding sensitivity to pitch. An important consequence is that these local embedding properties are directly reflected in nearest neighbor algorithms where, for two candidate tracks that are similar to an embedding query, the one that has a tempo very similar to the query is more likely to be selected. This may be undesirable in cases where the distance imposed by a difference in tempo or key outweighs the similarity of a property such as genre or mood, which may be more relevant to the query's intent.

To validate the change in sensitivity of embedding spaces to pitch and tempo features of unmodified audio, we consider the average variability of tempo, tags and key in the neighborhoods of each track over all test partitions of the AllTempo, MSD, and the entirety of the GS\textsubscript{Key} datasets. For tempo we collect the K-nearest neighbors in the cosine embedding space and calculate the root-mean-minimum-square (RMMS) distance between the closest tempo octave of $[1/3, 1/2, 1, 2, 3]$ of the seed track's tempo and each of its K neighbors. To investigate the locality of key in a nearest neighbor setting, we look at the precision of retrieving key labels from the K-nearest neighbors of each seed track relative to it's own tags. In all cases, metrics are averaged over all possible seeds in the dataset.

\begin{figure}[t!]
 \centerline{
 \includegraphics[trim={0 0.7cm 1.2cm 1.45cm},clip,width=1\columnwidth]{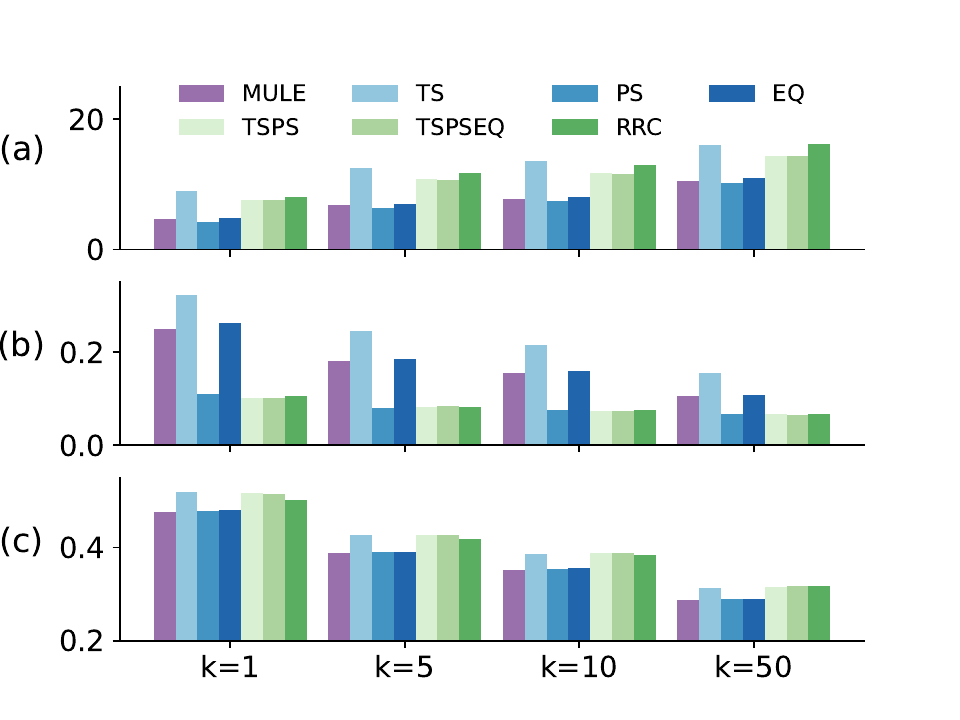}}
 \caption{Average metrics of neighborhoods of size $k$ for each fine-tuned model (a) Tempo RMMS distance (over AllTempo-test), (b) Key Precision (over GS\textsubscript{Key}), (c) Tag Precision (over MSD-test).}
 \label{fig:manipulated_tempo_neighborhood}
\end{figure}

The results of these experiments are shown in Fig.~\ref{fig:manipulated_tempo_neighborhood}. In Fig.~\ref{fig:manipulated_tempo_neighborhood}(a) we see that all augmentations involving time stretching do indeed increase the variability of tempo in local neighborhoods of the embedding space, whereas PS augmentation slightly reduces this. Intuitively this is as we expect, and agrees with Fig.~\ref{fig:manipulated_tempo}; if embeddings are less sensitive to one feature we expect them to be more sensitive to the remaining features due to reduced competition with the now-desensitized feature. Conversely, in Fig.~\ref{fig:manipulated_tempo_neighborhood}(b), we see that embeddings trained with pitch augmentation notably decrease the precision (and hence consistency) of neighboring embedding keys relative to a query embedding, whereas TS augmentation notably increases this precision. It is promising to see tag precision of neighboring embeddings increases when applying any form of time-stretching augmentation, highlighting this augmentation to be an important factor in improving the local organization of embedding spaces by tags such as genre and key. We note that RRC augmentation is not as effective at increasing the locality of tags in the embedding space. In all cases we observe EQ augmentation to have little effect.

\subsection{Nearest Neighbor Retrieval}
\label{subsec:nearestneighborretrieval}

The intention of embedding spaces that are desensitized to pitch, tempo and equalization is to improve the locality of properties that are relatively impartial to these, for downstream uses such as ANN retrieval. To evaluate, we follow the methodology in \cite{lee2020metric}, over the same MSD test partition. We compute a tag retrieval metric by evaluating for each tag, the percentage of tracks with that tag that have it in common with any of their k-nearest neighbors. This is averaged over all tracks and tags. We also compute the precision at k between each seed track's tag and the retrieved tags from its k-nearest neighbors, averaged over all seed tracks and tags. Results in Table~\ref{table:tag_retrieval} show MULE-based models outperform previous work, however, those tuned with TS augmentation perform significantly better than those without. Corroborating results in Section \ref{subsec:localembeddingproperties}, RRC augmentation improves the grouping of embeddings by tags in local neighborhoods, however, it is not as effective as TS augmentation.

\begin{table}[t!]
  \centering
  \footnotesize
  \begin{tabular}{@{}cccccccccccc@{}}
    \multirow{2}{*}{\textbf{}} & \multicolumn{3}{c}{Precision} & \multicolumn{4}{c}{Tag Retrieval} \\
     & $k$=2 & $k$=4 & $k$=8 & $k$=1 & $k$=2 & $k$=4 & $k$=8 \\
    \toprule[1.1pt]
    MULE & 44.1 & 40.1 & 36.3 & 47.6 & 59.7 & 71.0 & 80.7  \\
    \midrule
    TS & 48.0 & \textbf{44.1} & 39.9 & \textbf{51.9} & 63.3 & \textbf{73.9} & \textbf{82.5} \\
    PS & 44.1 & 40.2 & 36.6 & 47.8 & 59.6 & 71.1 & 80.7 \\
    EQ & 44.3 & 40.5 & 36.7 & 48.0 & 59.8 & 71.3 & 81.0 \\
    TSPS & \textbf{48.1} & \textbf{44.1} & 40.0 & 51.8 & 63.2 & 73.7 & 82.4 \\
    TSPSEQ & \textbf{48.1} & \textbf{44.1} & \textbf{40.1} & 51.6 & \textbf{63.4} & 73.8 & \textbf{82.5} \\
    RRC & 46.6 & 42.9 & 39.4 & 50.3 & 61.8 & 72.8 & 81.9 \\
    \midrule[1.1pt]
    \cite{lee2020metric}& - & - & - & 45.0 & 58.5 & 71.0 & 80.9 \\
    \bottomrule[1.1pt]
    \end{tabular}  
    \caption{Nearest neighbor content based retrieval performance for different augmentations. Metric definitions are equivalent for $k=1$.}
  \label{table:tag_retrieval}
\end{table}

\subsection{Music Labelling}
\label{sssec:subsubhead}

Here, we look at the performance of embeddings over pitch, tempo and other labeling tasks to both confirm the loss of pitch / tempo information in embeddings trained with PS / TS augmentation, and the retention of other information such as genre and instrumentation. 

For the AllTempo dataset we train a multi-layer perceptron as a 271 class classification problem for integer BPMs from 30 to 300~BPM. This probe has a single layer of 512 neurons and 75\% dropout, trained with a batch size of 256 over 20k~steps. Following \cite{bock2019multi}, the tempo estimate is taken as the \textit{argmax} of the classifier output after smoothing by a 15~BPM Hamming window. As there are no prior results on these datasets for generic music embeddings, we include two baselines: an end-to-end tempo convolutional network \cite{schreiber18ismir} and a SotA bespoke tempo and beat tracking model \cite{boeck20ismir}. Congruent with the results in Section~\ref{subsec:localembeddingproperties}, 
Table~\ref{table:nonmsd} shows TS augmentation degrades performance on tempo tasks, while PS augmentation improves it. The latter of which exceeds the performance of  \cite{schreiber18ismir}, approaching the performance of \cite{boeck20ismir}. This is promising considering that this baseline is highly task specific, whereas MULE is a generic music embedding that achieves excellent results in many other tasks.

For all other tasks we use the same probe training parameters as \cite{mccallum22ismir}. Table~\ref{table:labeling_tasks} shows that despite the heavy influence of tempo in the local organization of embedding spaces observed in Section~\ref{subsec:localembeddingproperties}, models with PS augmentation yield the greatest improvement on non-pitch dependent tasks (NSynth\textsubscript{I}, Gtzan and MTT). Of these tasks, combinations of augmentation strategies achieve SotA on MTT. Interestingly, while RRC augmentation degrades performance on MTT, we see SotA performance on the smaller Gtzan dataset.

\begin{table}[t!]
  \centering
  \footnotesize
  \begin{tabular}{@{}ccccccccccc@{}}
    \multirow{2}{*}{\textbf{}} & \multicolumn{2}{c}{Gtzan} & \multicolumn{2}{c}{Giantsteps} & \multicolumn{2}{c}{ACM-Mirum} \\
     & Acc1 & Acc2 & Acc1 & Acc2 & Acc1 & Acc2 \\
    \toprule[1.1pt]
    MULE & 74.1 & 90.5 & 85.5 & \textbf{98.2} & 81.2 & 95.8 \\
    \midrule
    TS & 54.6 & 66.2 & 71.0 & 84.1 & 66.2 & 76.9 \\
    PS & \textbf{76.2} & \textbf{91.1} & \textbf{86.7} & \textbf{98.2} & \textbf{82.3} & \textbf{96.4} \\
    EQ & 73.2 & 90.2 & 77.0 & 98.0 & 81.1 & 95.6 \\
    TSPS & 64.1 & 78.0 & 78.8 & 92.3 & 76.5 & 87.7 \\
    TSPSEQ & 60.4 & 75.5 & 56.9 & 75.0 & 68.7 & 79.9 \\
    RRC & 54.0 & 67.1 & 49.2 & 69.7 & 63.0 & 73.7 \\
    \midrule[1.1pt]
    \cite{schreiber18ismir} & 76.9 & 92.6 & 82.1 & 97.1 & 78.1 & 97.6 \\
    \cite{boeck20ismir}& \textbf{83.0} & \textbf{95.0} & \textbf{87.0} & 96.5 & \textbf{84.1} & \textbf{99.0} \\
    \bottomrule[1.1pt]
    \end{tabular}  
    \caption{Acc 1 and Acc 2 metrics \cite{gouyon06taslp} on AllTempo test partitions.}
  \label{table:nonmsd}
\end{table}

\begin{table}[t!]
  \centering
  \footnotesize
  \begin{tabular}{@{}ccccccccccc@{}}
    \multirow{2}{*}{\textbf{}} & GS\textsubscript{Key} & NSynth\textsubscript{P} & NSnyth\textsubscript{I} & Gtzan & \multicolumn{2}{c}{MTT} \\
     & W. Acc & Acc & Acc & Acc & mAP & ROC \\
    \toprule[1.1pt]
    MULE & 66.7 & 89.2 & 74.0 & 73.5 & 40.4 & 91.4 \\
    \midrule
    TS & \textbf{67.2} & 89.0 & 74.8 & 77.2 & 40.4 & 91.5 \\
    PS & 30.2 & 81.2 & 75.0 & 79.7 & 40.5 & 91.5 \\
    EQ & 65.9 & \textbf{89.5} & 73.6 & 75.5 & 40.5 & 91.4 \\
    TSPS & 18.5 & 79.3 & 73.2 & 81.0 & 40.8 & \textbf{91.7} \\
    TSPSEQ & 17.6 & 79.7 & \textbf{75.3} & 80.3 & \textbf{40.9} & \textbf{91.7} \\
    RRC & 14.9 & 76.9 & 73.9 & \textbf{82.8} & 37.1 & 89.2 \\
    \midrule[1.1pt]
    \multirow{2}{*}{SS SotA} & \textbf{67.3} & \textbf{94.4} & \textbf{78.2} & 81.1 & \textbf{40.9} & 91.4 \\
     & \cite{li2023mert} & \cite{li2023mert} & \cite{wang2022towards} & \cite{zhao2022s3t} & \cite{zhao2022s3t} & \cite{mccallum22ismir} \\
    \bottomrule[1.1pt]
    \end{tabular}  
    \caption{Tagging performance of augmentation pipelines with compared to no augmenation (MULE) and self-supervised (SS) SotA.}
  \label{table:labeling_tasks}
\end{table}

\section{Conclusions}
\label{sec:conclusion}

In this work we investigated intra-pair data augmentation on the local properties of contrastive music audio embedding spaces and showed its effect on downstream nearest neighbor / labelling tasks, with several key results. Firstly, we demonstrate that local neighborhoods in contrastively learned audio embedding spaces reflect local properties in the training data. Secondly, applying data augmentation mitigates the locality of related properties in the embedding space, while improving the locality of unrelated properties. Thirdly, the optimal selection of data augmentation strategies is task dependent, and careful selection results in SotA performance on downstream labelling / nearest neighbor tasks. Finally, we identify tempo as a key feature in the organization of contrastive music audio embedding spaces.

\bibliographystyle{IEEEbib2}
\bibliography{article}

\end{document}